\documentclass[12pt,
pre,
aps,
superscriptaddress,
footinbib,
showkeys,
titlepage
]{revtex4-1}

\def\papertitle{Superionic liquids in conducting nanoslits: A variety of phase transitions and ensuing charging behavior}

\usepackage{xspace}

\usepackage[draft,inline,nomargin]{fixme}
\fxsetup{theme=color,mode=multiuser}
\FXRegisterAuthor{sk}{ask}{\color{red}SK}

\usepackage{amssymb}
\usepackage{amsmath}
\usepackage{commath}
\usepackage{graphicx,bm}
\usepackage[dvipsnames]{xcolor}

\usepackage[utf8]{inputenc}

\usepackage{textcomp}
\DeclareUnicodeCharacter{2212}{-}

\newlength{\figwidth}

\newif\ifOnecolumn
  \ifx\Onecolumn\undefined
  \Onecolumntrue
  
\else
  \Onecolumnfalse
\fi

\newif\ifTwocolumn
  \ifOnecolumn
   \Twocolumnfalse
\else
  \Twocolumntrue
\fi

\newcommand{\latin}[1]{{\it #1}}
\newcommand{\ie}{\latin{i.e.}\@\xspace}
\newcommand{\eg}{\latin{e.g.}\@\xspace}
\newcommand{\cf}{\latin{cf.}\@\xspace}
\newcommand{\etc}{\latin{etc.}\@\xspace}
\newcommand{\viz}{\latin{viz.}\@\xspace}

\def\ba{\begin{align}}
\def\ea{\end{align}}

\newcommand{\fig}[1]{Fig.~\ref{#1}}
\newcommand{\figs}[1]{Figs.~\ref{#1}}

\newcommand{\Fig}[1]{Figure~\ref{#1}}

\newcommand{\app}[1]{Appendix~\ref{#1}\@\xspace}
\newcommand{\sect}[1]{Section~\ref{#1}\@\xspace}

\newcommand{\eq}[1]{Eq.~(\ref{#1})}
\newcommand{\eqs}[2]{Eqs.~\eqref{#1} and \eqref{#2}}
\newcommand{\Eq}[1]{Equation~(\ref{#1})}
\newcommand{\mycite}[1]{Ref.~\cite{#1}}

\newcommand{\e}{\mathrm{e}}

\begin{document}

\title{\papertitle}

\author{Maxym Dudka}
\affiliation{Institute for Condensed Matter Physics of the National Academy of Sciences of Ukraine, 1
Svientsitskii st., 79011 Lviv, Ukraine}
\affiliation{${\mathbb L}^4$ Collaboration \& Doctoral College for the Statistical Physics of Complex Systems, Leipzig-Lorraine-Lviv-Coventry, Europe }
\email{maxdudka@gmail.com}

\author{Svyatoslav Kondrat}
\affiliation{Department of Complex Systems, Institute of Physical Chemistry PAS, Kasprzaka 44/52, Warsaw, Poland}
\email{skondrat@ichf.edu.pl; svyatoslav.kondrat@gmail.com}

\author{Olivier B\'enichou}
\affiliation{Sorbonne Universit\'e, CNRS, Laboratoire de Physique Th\'eorique de la Mati\`{e}re Condens\'ee, LPTMC (UMR CNRS 7600), 75252 Paris Cedex 05, France}

\author{Alexei A. Kornyshev}
\affiliation{Department of Chemistry,  Molecular Sciences Research Hub, White City Campus, London W12 0BZ,United Kingdom}
\affiliation{Thomas Young Centre for Theory and Simulation of Materials, Imperial College London, South Kensington Campus, London SW7 2AZ, United Kingdom}
\email{a.kornyshev@imperial.ac.uk}

\author{Gleb Oshanin}
\affiliation{Sorbonne Universit\'e, CNRS, Laboratoire de Physique Th\'eorique de la Mati\`{e}re Condens\'ee, LPTMC (UMR CNRS 7600), 75252 Paris Cedex 05, France}
\affiliation{Interdisciplinary Scientific Center J-V Poncelet, UMI CNRS 2615, 11 Bol. Vlassievsky per., Moscow, Russia}
\email{oshanin@lptl.jussieu.fr}

\begin{abstract}

    We develop a theory of charge storage in ultra-narrow slit-like pores of nano\-structured electrodes. Our analysis is based on the Blume-Capel model in external field, which we solve analytically on a Bethe lattice. The obtained solutions allow us to explore the complete phase diagram of confined ionic liquids in terms of the key parameters characterising the system, such as pore ionophilicity, interionic interaction energy and voltage. The phase diagram includes the lines of first and second-order, direct and re-entrant, phase transitions, which are manifested by singularities in the corresponding capacitance-voltage plots. To test our predictions experimentally requires mono-disperse, conducting, ultra-narrow slit pores, permitting only one layer of ions, and thick pore walls, preventing interionic interactions across the pore walls. However, some qualitative features, which distinguish the behavior of ionophilic and ionophobic pores, and its underlying physics, may emerge in future experimental studies of more complex electrode structures.

\end{abstract}

\maketitle 

\section{Introduction}

    Supercapacitors store energy via charge separation between cathode and anode in electrical double layers at the corresponding electrodes \cite{conway:99}. Since the stored energy scales up with the electrode/electrolyte contact area, highly porous electrodes with 'volume-filling' surfaces are used to improve their capacitive performance. If the pores of an electrode are much wider than the double layer thickness, the capacitance \emph{per surface area}, provided by this electrode, will be comparable to the corresponding capacitance of a flat electrode \cite{Daikhin1996}.  However, if the double layers on the opposite sides of the pores overlap, or if the pores are so narrow that they can accommodate only one layer of ions, then the physics and the laws of charge storage become different \cite{gogotsi:sci:06, pinero:carbon:06, gogotsi:08, kondrat:jpcm:11}. The need to maximize the capacitance and energy storage motivates the use of electrodes with such fine porosity, and, since the modern nanotechnology allows one to build well-defined nanoporous structures \cite{liu:nanolett:10, zhu, tsai:ne:13, chen:jpcl:graphene:13, lukatskaya13a, Naguib2013, Zhao2014}, understanding the laws of charge storage in nano-sized pores has become in great demand \cite{kondrat:jpcm:11, feng:jpcl:11, merlet:natmat:12, vatamanu:jpcl:energystorage:13, vatamanu:acsnano:15, Ma2017, Mendez-Morales2018, Breitsprecher2018, Mossa2018}.

While in computer simulations the pores of various types \cite{vatamanu:acsnano:15, Ma2017, Mendez-Morales2018}, or even nanoporous networks \cite{merlet:natmat:12}, are modeled in a conceptually similar way \cite{Mendez-Morales2019}, analytical theories of ultra-nanoporous electrodes, comparable in size to the size of a bare ion, require special modeling, which can, however, benefit from the `reduced dimensionality' of the system. For instance, a narrow slit nanopore may be treated as a quasi two-dimensional system of ions, which are drawn into the pore by electrode polarization; for a single-file cylindrical pore, the system is even quasi one-dimensional (1D). 

Such quasi-1D systems have been studied by mapping them on the corresponding models of statistical mechanics \cite{Horgan1DIsingCaoacitor:2012, Horgan1DIsingOverscreening:2012, Demery2016, schmickler:ea:2015:harmonicOscillator, lee:nanotech:14, kornyshev:fd:14, lee:prl:14, rochester:jpcc:16, Frydel2018}. The simplest appropriate model is a classical two-state anti-ferromagnetic Ising model with nearest neighbor interactions in external field \cite{kornyshev:fd:14}. In this model, $\pm 1$ spins correspond to positive and negative ions, \ie, cations and anions, respectively, and the external magnetic field is the potential drop between an electrode and bulk electrolyte; the approximation of nearest neighbor interactions is reasonable due to the superionic state emerging in conducting nanoconfinement \cite{kondrat:jpcm:11}, in which the inter-ionic interactions are exponentially screened \cite{rochester:cpc:13}. The solution to this model is well known and can be found in textbooks \cite{Baxter1982}, but it allowed a compact analytical expression for the capacitance and the in-depth analysis of its behavior \cite{kornyshev:fd:14}. Despite its simplicity, this model and its extensions have turned out to capture relatively well the qualitative behavior of the voltage-dependent capacitance, as compared to simulations \cite{lee:prl:14, rochester:jpcc:16}.

    While for 1D models, describing single-file cylindrical pores, there are exact analytical solutions, the analytical solutions to 2D (and higher dimension) problems are limited, particularly in `external field', the role of which is played by the applied potential. It seems therefore rewarding to resort to approximate but reliable approaches, which provide analytical solutions and hence allow new physical insights to be more easily developed. In this work, we present a three-component model for ionic liquids (and solvent, or voids) confined into narrow slits, which are subject to externally applied voltage. We solve this model \emph{exactly} on a Bethe lattice \footnotemark[1] with coordination number $q$, which approximate lattices with the same coordination number (it can also be seen as an approximation to off-lattice systems with, on average, $q$ neighbors) \footnotemark[2]. This model can be mapped onto the well known antiferromagnetic Blume-Capel model, which has been treated by the Bethe-lattice approach \cite{Ekiz2004}, cluster-variation method \cite{Rosengren1993} and simulations~\cite{Kimel86, Kimel92, Wilding1996, Pawlowski06, Zukovic2013} in the context of magnetic systems. However, the aspects of the phase diagram related to the response to electric field, as well as charging and capacitive characteristics, relevant to confined ionic liquids and supercapacitors, have not been investigated. 

\footnotetext[1]{The Bethe lattice is a deep interior part of the infinite graph called the Cayley tree, see \mycite{Gujrati95} for details.}
\footnotetext[2]{We have chosen the coordination number $q=3$, corresponding to a hexagonal lattice, but we do not expect any qualitative changes in the system behavior for other values of $q$. We note in particular that in \mycite{dudka:jpcm:16} the coordination numbers $q=3$ and $q=4$ have been considered for the case of zero applied voltage; while the locations of the transitions were $q$-dependent, qualitatively the same phase behavior was observed.}

    In our previous work \cite{dudka:jpcm:16}, we have applied this model to analyse the phase behavior of ions in \emph{non-polarized} slit nanoconfinement. We demonstrated the emergence of ordered and disordered phases, and first and second-order phase transitions, which could be induced by changing temperature and slit width. Herein, we extend this analysis to a much more complex case of polarizable electrodes and study the \emph{voltage-dependent} phase behavior and how it projects onto the charging properties. The introduction of the electrode potential is not only necessary to study the capacitive characteristics, but we shall see that it also leads to a far richer phase behavior with a variety of direct and re-entrant phase transitions, giving rise to remarkable charging properties. Since the ordered phases are known to exhibit slow dynamics \cite{kondrat:nm:14, he15a, Breitsprecher2018}, our work may also have an important practical implication in that the knowledge of the parameter space corresponding to such phases may help avoid potential slowdown of charging.

 The paper is organized as follows: The model is formulated in \sect{sec:model}, in which we also discuss its shortcomings and possible extensions (\sect{sec:limits}); the Bethe-lattice solution to this model is presented in \app{app:calcs}. The results are discussed in \sect{sec:res}, and we summarize and conclude in \sect{sec:concl}. 

\section{Model}
\label{sec:model}

\begin{figure}[t]
\begin{center}
	\includegraphics[width=0.6\textwidth]{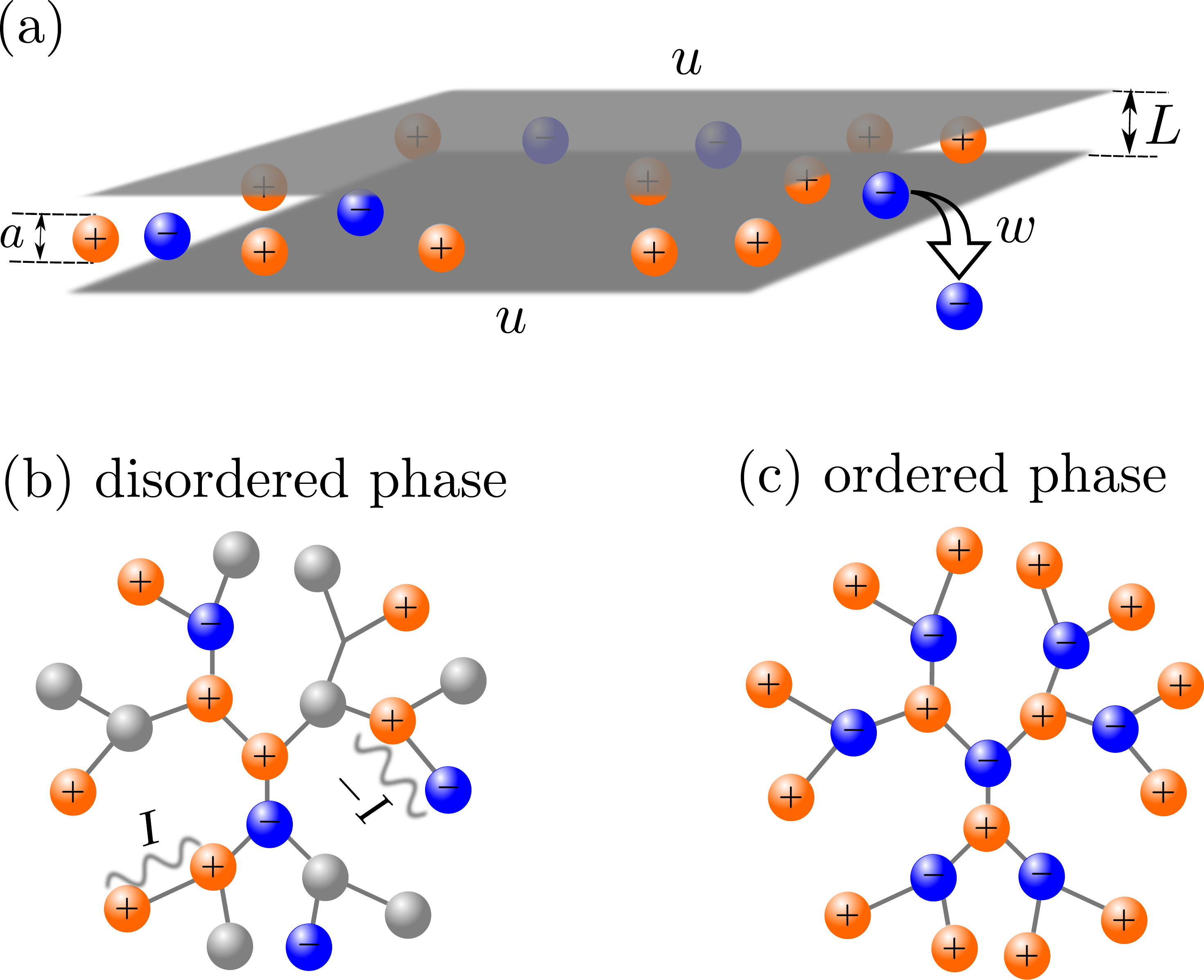}
    \caption{Model of ions inside a slit nanopore and the Cayley tree representation. (a) Ions are confined into a narrow metallic slit of width $L$. Ion diameter is $a=a_\pm$, which is assumed the same for cations and anions. Potential $u$ is applied at the pore walls with respect to the bulk electrolyte (not shown). The energy of transfer of ions from a pore into the bulk is $w=w_\pm$ and is assumed the same for both ions (see text). (b) Fragment of the Cayley tree with coordination number $q=3$, illustrating a possible arrangement of ions (blue and orange spheres) and voids/solvent (gray spheres) in the disordered phase. Interaction energy between the same type of ions is $I$ and between oppositely charged ions is $-I$ (see \eq{eq:H}). (c) The same fragment of the Cayley tree as in (a), but showing an example of the ordered phase in the case of high transfer energy ($\beta w \gg 1$) and zero applied potential ($u=0$); in this case, the ordered phase consists of an equal amount of cations and anions occupying two sublattices of the Cayley tree.
    \label{fig:model}
}
\end{center}
\end{figure}

We consider an ultra-narrow metallic slit-shaped pore, just one molecular diameter thick, such that it can accommodate only one layer of ions (\fig{fig:model}a; for simplicity, we assume the cations and anions to be of comparable size). We also assume that the ions inside the pore reside on the sites of a regular lattice with coordination number $q$ (we note that lattice models are frequently used for ionic liquids and seem to capture their behavior qualitatively well \cite{kornyshev:07, Horgan1DIsingCaoacitor:2012, Horgan1DIsingOverscreening:2012, Demery2016, lee:nanotech:14, kornyshev:fd:14, lee:prl:14, rochester:jpcc:16, dudka:jpcm:16, Girotto2018, Girotto2017a}).  The lattice is thus populated by a mixture of cations, anions and voids (empty, unoccupied sites). To describe the occupation of site $i$, it is convenient to introduce Boolean variables $n^\pm_i = 0$ or $1$, such that $n^+_i = 1$ ($n^-_i = 1$) means that site $i$ is occupied by a cation (an anion), and $n^+_i= n^-_i = 0$ means that site $i$ is vacant; the configuration $n^+_i = n^-_i = 1$ is prohibited due to hard-core exclusion (\ie, each site can be occupied either by a cation or an anion, or be empty). This system can be described by the following Hamiltonian \cite{dudka:jpcm:16}
\begin{align} 
\label{eq:H} 
{\cal H} = I \sum_{\langle i,j\rangle} \big(n^+_i n^+_j + n^-_i n^-_j  - n^+_i n^-_j - n^-_i n^+_j \big) - \sum_i \big(h_- n^-_i + h_+ n^+_i\big), 
\end{align}
where $\langle ij \rangle$ denotes nearest neighbor sites, $I > 0$ is the interaction energy between two \emph{neighboring} ions, and the interactions between the next-nearest and further neighbors are neglected (this is seemingly an acceptable approximation due to the superionic state, \ie an exponential screening of ionic interactions, emerging in conducting nano-confinement \cite{kondrat:jpcm:11}; see, however, \sect{sec:limits:nnn}). In \eq{eq:H}, `external' fields $h_{\pm}$ are  
\begin{align}
    \label{eq:h}
	h_\pm = w_\pm \pm e u,
\end{align}
where $u$ is the applied potential (with respect to the bulk electrolyte outside of the pore), $e$ is the elementary charge and $w_\pm$ is the energy of transfer of a $\pm$ ion \emph{from} the pore into the bulk, which includes the image forces and other interactions of the ions with the pore walls \cite{kondrat:jpcm:11}. We note that the definition used here differs by sign from the re-solvation energy used in other works~\cite{kondrat:jpcm:11, kondrat:pccp:11, kondrat:ec:13} (here, positive $w$ corresponds to ionophilic and negative to ionophobic pores). In the following we assume $w_+ = w_- = w$ and note that the results for an asymmetric ionic liquid ($w_+ \neq w_-$) can be obtained by shifting the applied voltage by $-(w_+ - w_-)/2e$ and taking the transfer energy equal to $(w_+ + w_-)/2$.

Thermodynamic properties can be calculated from the partition function 
\begin{align} 
\label{eq:partition} 
\Xi = \sum_{(n^+_i,n^-_i)} \exp\{-\beta \mathcal{H}\} \,, 
\end{align} 
where $\beta = 1/k_B T$, $k_B$ is the Boltzmann constant and $T$ temperature, and the sum runs over all possible allowed configurations $(n^+_i,n^-_i)$ of the occupation variables. We have calculated the partition function \eqref{eq:partition} by using the Bethe-lattice approach, which relies on approximating the actual (lattice) structure by a Bethe lattice \footnotemark[1] with the same coordination number \footnotemark[2] (\fig{fig:model}b,c); the partition function on the Bethe lattice can then be evaluated exactly  \footnote[3]{See recent \mycite{dudka:jpcm:16, Dudka2018} and references therein for a detailed description of the Bethe lattice approach} (see \app{app:calcs} for details).

\subsection{Charging characteristics}

Having calculated the partition function \eqref{eq:partition}, one can compute the charge accumulated in a pore
 \begin{align}
     \label{eq:Q}
	Q = - \frac{1}\beta \frac{d \ln(\Xi) }{d u},
\end{align} 
and the differential capacitance
\begin{align}
    \label{eq:cap}
	C = \frac{d Q}{d u}.
\end{align}
Note that per definition both $Q$ and $C$ are measured per surface area (since we consider a quasi two-dimensional problem); both quantities can be assessed experimentally~\cite{fedorov_kornyshev:chemrev:14}. 

We shall also calculate the charging parameter~\cite{forse:jacs:16:chmec, breitsprecher:jcp:17:mcmd}
\begin{align}
	\label{eq:XD}
	X_D = \frac{e}{C} \frac{d \rho}{d u} = e \frac{d\rho}{d Q}, 
\end{align} 
where $\rho(u)$ is the total (two-dimensional) ion density in a pore. This parameter describes charging mechanisms taking place at given applied voltage $u$. In particular, $X_D = 0$ means that the in-pore total ion density does not change with $u$, which implies that charging, in the thermodynamic sense, is driven by swapping the in-pore co-ions for the counter-ions from the reservoir (bulk electrolyte). $X_D = 1$ indicates that charging proceeds by electorosorption of new counter-ions from bulk electrolyte, implying that the in-pore total ion density increases along with the charge. $X_D=-1$ means that charging is due to desorption of the in-pore co-ions and hence the total ion density decreases as the charge increases. For $|X_D| < 1$, charging is a combination of swapping and adsorption/desorption, with the contribution from swapping $1 - |X_D|$. It is also possible that $X_D > 1$, which means that, in addition to counter-ions, also co-ions are adsorbed into a pore; similarly, $X_D < -1$ implies that, alongside the in-pore co-ions, also the counter-ions are removed from a pore.

\subsection{Ordered and disordered phases}
\label{sec:phases}

By using the appropriate change of variables and of the interaction parameters (see, \eg, \mycite{dudka:jpcm:16}), the lattice model given by \eq{eq:H} can be mapped onto the classical spin-$1$ Blume-Capel (BC) model. The BC model has been studied by cluster-variation \cite{Rosengren1993} and Bethe-lattice approaches \cite{Ekiz2004} and by Monte Carlo simulations~\cite{Kimel86, Kimel92, Wilding1996, Pawlowski06, Zukovic2013}. Based on these results, we expect rich phase behavior, involving direct and re-entrant symmetry-breaking phase transitions between `ordered' and `disordered' phases. In the context of ionic liquids, the disordered phase is a homogeneous mixture of ions and voids (\fig{fig:model}b), while the `ordered' phase means that the ions of one type predominantly occupy one of the two `sublattices'. At zero applied potential, $u=0$, the ordered phase consists of an equal amount of cations and anions (possibly mixed with voids, depending on the transfer energy $w$, see \fig{fig:model}c for the case $\beta w \gg 1$); in the corresponding continuous system, this likely corresponds to in-plane crystal-like ordering of ions \cite{dudka:jpcm:16}. At a non zero potential and in the ordered phase, the charge on one sublattice exceeds that on the other, so that the pore is overall charged. A particular example is an ordered phase consisting of counter-ions and voids, achievable at sufficiently high potentials. Other ordered configurations are also possible within the underlying lattice structure.

\subsection{Discussion of model limitations}
\label{sec:limits}

As mentioned, we have solved model \eqref{eq:H} on the Bethe lattice, which allowed us to calculate various charging characteristics \emph{analytically}. Before describing the results of these calculations, however, it is important to discuss the limitations of the model, its possible extensions and the relation to charging real nanoporous electrodes.

\subsubsection{Next-nearest and further neighbors}
\label{sec:limits:nnn}

We have assumed that only the nearest-neighbor ions interact with each others. This assumption has been made because of the superionic state emerging in conducting nanoconfinement. Indeed, for perfectly metallic walls of a slit pore, the interaction energy between two ions, situated in the slit's midplane and separated by distance $r$, is $U_{\alpha\beta}(r) = z_\alpha z_\beta \phi(r)$, where \cite{kondrat:jpcm:11} 
\begin{align}
    \label{eq:phi}
	\phi (r) =  \frac{4 e^2}{\varepsilon L} \sum_{n=1}^\infty  K_0 \left( \pi n r/L\right).
\end{align}
Here $K_0 (x)$ is the modified Bessel  function  of  the  second  kind  of  order zero; $\alpha, \beta=\{+,-\} $, $z_\alpha$ is the ion charge, measured in units of the elementary charge $e$, $L$ is the pore width, and $\varepsilon$ is the dielectric constant inside the pore (we use Gaussian units throughout the paper). For this system and monovalent ions ($z_\pm=\pm 1$), the coupling constant in \eq{eq:H} is $I = \phi (a)$, where $a$ is the lattice constant. In this case, the value of $I$ varies in the range from a few $k_BT$ to about $15 k_BT$, depending on the ion size ($\approx$ lattice constant), pore width and dielectric constant \cite{dudka:jpcm:16}.

For large separations, $r \gg L$, the large-argument asymptotic expression for the Bessel function \cite{gradstein:81} gives
\begin{align}
	\phi (r) =  \frac{4 e^2}{ \varepsilon \sqrt{2L r}} \e^{- \pi r/L},
\end{align}
\ie, $\phi(r)$ decays exponentially with $r$. For the lattice constant $a=1$nm and slit width $L=a$, and taking the dielectric constant $\varepsilon = 2$ (see \sect{sec:limits:epsilon}), one finds $I = \phi(a) \approx 3.3 k_BT$, while $\phi (2a) \approx 0.1 k_BT$ (both at room temperature); thus, the interaction between the next-nearest-neighboring ions is $\alpha = \phi(a)/\phi(2a) \approx 33$ times smaller than the interaction between the neighboring ions. This seemingly justifies the approximation made above. However, if the pore widths is just below two ion diameters, the decay length becomes two times larger, and hence $\alpha \approx 7$; then the next-nearest interactions are not necessarily so negligible any more. \citeauthor{Badehdah1998} have demonstrated that inclusion of the second neighbors can lead to the appearance of new states and multicritical points \cite{Badehdah1998}. It would be interesting to investigate such effects in future work. 

\subsubsection{Ion polarizability}
\label{sec:limits:epsilon}

It is known that ion polarizability can play an important role in formation of electrical double layes \cite{ Schroeder2010, Frydel2011, Bordin2016, Bedrov2019}, and its \textit{bona fide} modelling can be essential, particularly at high ion concentrations and for low dielectric media \cite{Frydel2011}. In our classical statistical-mechanical model, the conformational and electronic degrees of freedom (of ions), responsible for ion polarizability, are not considered explicitly, but enter the model only through the effective dielectric constant $\varepsilon$ of the interior of a pore, see \eq{eq:phi}. For pores densely packed with ions, and in the absence of solvent, we expect $\varepsilon$ to be in the range between $2$ and $5$, depending on the slit width; however, the dielectric constant presumably decreases and approaches unity as the in-pore ion density decreases \cite{kondrat:ec:13}. In other words, the dielectric constant in \eq{eq:phi} depends on the total ion density, which leads to complicated transcendental equations for ion densities. While such changes in the ion polarizability may affect the location (and the order) of the phase transitions discussed in this work, we do not expect the transitions to disappear and anticipate a similar qualitative behavior. (However, it is interesting to note, in this respect, that the pore-width dependence of the dielectric constant has been shown to have a potentially vivid effect on the charging behavior \cite{kondrat:ec:13}.) 

\subsubsection{Presence of solvent}

It might be tempting to interpret vacant sites (voids) as solvent molecules, by appropriately adjusting the effective dielectric constant inside a pore, particularly for systems with non-polar or weakly polar solvents, comparable in size to ions. This has been done in the previous work on charging cylindrical pores \cite{lee:prl:14, rochester:jpcc:16}, and even the solvent/ion size asymmetry was taken into account \cite{rochester:jpcc:16}. Although similar reasoning applies also to our model, we note that such interpretations shall be made with caution. Indeed, in real systems there are both `voids' \emph{and} solvent present, and the ions may interact with solvent not only by excluded volume interactions. For instance, it has been shown that polar solvents may affect charging significantly, leading to an increase in the capacitance, which is not captured by models with the ions and solvent interacting sterically only \cite{Jiang2014}. Thus, while our model may be applicable in the case of some solvents, it would be really interesting to extend it to treat solvent more rigorously.

\subsubsection{On the lattice nature of the model and Bethe-lattice approach}

Model \eqref{eq:H} is formulated on a bipartite lattice with coordination number $q$ (the number of nearest neighbors; $q=3$ in this work). This allowed us to define unambiguously the order parameter, \viz, the difference in the ion densities on two sublattices, and hence determine the locations of various phase transitions. In reality, however, the ions are not confined to reside on lattice sites, particularly in the disordered phase, and, upon transition to an ordered phase, may adopt a structure different from the prescribed lattice.

However, to solve our model, \ie, to calculate the partition function \eqref{eq:partition}, we have used the Bethe-lattice approach, which is based on the reformulation of the original lattice model onto a Bethe lattice (\fig{fig:model}b,c); the only information contained in the Bethe lattice about the lattice structure is the number of nearest neighbors. In this respect, our model may also be considered as an approximation to off-lattice systems, in which ions have $q$ neighbors separated by distance $a$  (on average). We emphasize that the partition function, and hence the ion densities and the capacitance, as calculated on the Bethe lattice, are \emph{exact}. It seems thus reasonable to expect that our analytical results reflect properly some physics of real ionic liquids in conducting slit nano-confinements.

\subsubsection{Carbonic vs metallic pore walls}

We have assumed that the pore walls are perfectly metallic surfaces, but this is not generally the case, as the majority of porous electrodes are fabricated from carbon materials \cite{Simon2008a}. This has three important implications in the context of our model.

\begin{enumerate}

    \item The screening of ionic interactions, \eq{eq:phi}, is expected to be different for carbonic pore walls \cite{rochester:cpc:13}. However, it has been shown by quantum density functional calculations that a similar expression for metallic cylinder is in fact a good approximation for single-wall carbon nanotubes \cite{Goduljan2014a, Mohammadzadeh2015a}. Although such calculations have not been performed for slits, it is reasonable to expect a similar agreement also in this case. It must be noted, however, that, depending on the pore-wall thickness, screening might be weaker for carbonic walls, as compared to the metallic one. Thus, the interaction energy between the next-nearest (and higher) neighboring ions may increase, which may produce a substantial effect on the phase behavior and capacitance (see \sect{sec:limits:nnn}).

    \item There is a contribution to the total (measurable) capacitance $C_\mathrm{tot}^{-1} = C_\mathrm{QC}^{-1} + C_\mathrm{IL}^{-1}$ from quantum capacitance $C_\mathrm{QC}$ of carbonic walls \cite{Xia2009, Kornyshev2013} ($C_\mathrm{QC} = \infty$ for metallic walls; $C_\mathrm{IL} \equiv C$ is the capacitance of an ionic liquid calculated in this work, \eq{eq:cap}). Thus, the capacitance-voltage curves, as computed in this work, will be modified by $C_\mathrm{QC}$, which is also voltage-dependent. However, since the phase transitions, discussed here, are manifested by strong peaks or divergencies in $C_\mathrm{IL}$ (\sect{sec:res}), they shall in principle be present in the total capacitance; it will be beneficial to account for $C_\mathrm{QC}$ explicitly in future work.

    \item Thin carbonic pore walls can be `transparent' to electrostatic (and other) interactions. Indeed, \citeauthor{Mendez-Morales2018} modelled a pore wall as a single layer of Gaussian charges and observed that the ions from neighboring pores are strongly correlated \cite{Mendez-Morales2018}. \citeauthor{Juarez2018} \cite{Juarez2018} used quantum density functional calculations and reported on similar correlations between the ions from the inside and outside of a single-wall carbon nanotube \cite{Juarez2018}. Most recently, it has been shown that such interactions may have a profound effect on the capacitance and energy storage \cite{Kondrat2019a}. In this work, we neglect the inter-pore ionic interactions, but it would be interesting to study how they influence the phase transitions discussed below; we note, however, that such effects shall be negligible for sufficiently thick pore walls.

\end{enumerate}

\subsubsection{Nanoporous electrodes}

Frequently used nanoporous electrodes for supercapacitors are activated and carbide-derived carbons \cite{Simon2008a}. Such electrodes do not consist of perfectly aligned monodisperse slits, but they are random porous media, containing interconnected pores of various shapes and sizes  \cite{Toso2013, prehal:17:pccp, prehal:17:nenergy, Vasilyev2019a, Prehal2019}. Clearly, our model is not directly applicable to these electrodes.

However, recently there has been a significant progress in developing low-dimensional carbon materials, such as quasi two-dimensional MXene phases \cite{lukatskaya13a, Naguib2013, Zhao2014} and graphenes \cite{liu:nanolett:10, zhu, tsai:ne:13, chen:jpcl:graphene:13}, which appear to be more suitable to test the predictions of our theory. It is important to note, however, that, even in these materials, the pore walls may not be perfectly aligned, the pore width may vary along a pore, and the walls can be rough or contain impurities. This will affect the strength of inter-ionic interactions locally and may allow multilayer filling of the pore with ions; all this may lead to rounding-off of the transitions, discussed in this work, or to turning them into smooth transformations between `ordered' and disordered phases. Nevertheless, it seems reasonable to expect that some features of these transitions, such as strong peaks manifested in the differential capacitance (\sect{sec:res}), shall still be present in the capacitance curves.

\subsubsection{Concluding remark on the model}

As discussed, the presented model, \eq{eq:H}, neglects many aspects of charge storage in nanoporous electrodes, and its predictions may not be easily verified in experiments due to the enormous complexity of real nanoporous electrode/electrolyte systems. Nevertheless, this model lends itself to be the \emph{simplest analytically-solvable model} for quasi-2D slit pores, likely capturing the basic physics of ionic liquids in ultra-narrow conducting slits, and thereby providing a reference frame for future theoretical, experimental and numerical analysis.

\section{Results and discussion}
\label{sec:res}

\begin{figure}[ht]
\begin{center}
	\includegraphics[width=\textwidth]{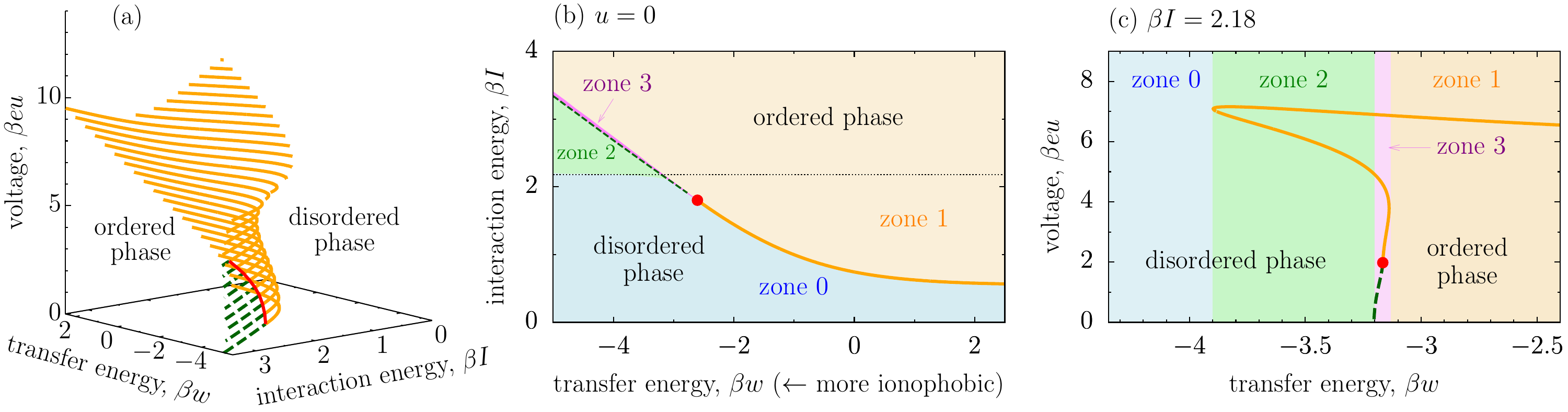}
    \caption{Phase diagram of a model superionic liquid described by \eq{eq:H}. (a) Phase diagram drawn in the ($w$, $I$, $u$) space, where $w$ is the ion transfer energy, $I$ the ion-ion interaction parameter, and $u$ the applied voltage. Ordered and disordered phases are separated by surfaces of first-order (dash green lines) and second-order (solid orange lines) phase transitions, which meet at a line of tricritical points (thick solid red line). At high voltages, the surface of second-order transitions (solid orange lines) bends down such that the disordered phase is always on top of the ordered phase, \cf panel (c). (b) Phase diagram at zero applied voltage ($u=0$). The red filled circle denotes a tricritical point. Differently coloured zones are numbered according to the number of phase transitions, which the systems belonging to these zone experience, as the voltage increases. The thin horizontal line shows the value of $\beta I$ used in (c). (c) Phase diagram in the ($w$, $u$) plane for $\beta I = 2.18$ (which lies close to the border line between zones $0$ and $2$). Voltage is measured in units of thermal voltage $(\beta e)^{-1}$ ($\approx 26$mV at room temperature).
\label{fig:pd}
}
\end{center}
\end{figure}

\Fig{fig:pd}a shows the global phase diagram, drawn in the space of ion's transfer energy $w$, ion-ion interaction energy $I$, and applied potential $u$. It consists of two phases, ordered and disordered (\sect{sec:phases}, \fig{fig:model}b,c), which are separated by two non-overlapping surfaces of first and second-order phase transitions. These surfaces meet at a line of tricritical points. It is important to note that the surface of second-order transitions bends down for increasing $u$, such that the disordered phase is always on top of the ordered phase (\cf also \fig{fig:pd}c). This means that, at sufficiently high applied potentials, the disordered phase becomes stable, independently of the values of $w$ and $I$.

To better understand the topology of the global phase diagram, we combined the phase diagram at zero potential, $u=0$, with  the \emph{orthogonal projection} of the phase transition surfaces onto the $u=0$ plane (\fig{fig:pd}b). This projection divides the $u=0$ plane into two basic regions: 
\begin{enumerate}

    \item The region onto which the transition surface does not project (zone 0 in \fig{fig:pd}b). A system with $I$ and $w$ belonging to this zone does not undergo any voltage-induced phase transition. In other words, the systems is, and remains, in the disordered state at any applied potential.

    \item The region onto which the transition surface projects at least once (if the surface bends or makes zigzags, it can projects twice or more, \cf \fig{fig:pd}c). This region is most interesting as it contains a number of voltage-induced phase transitions. Depending on the number of these transitions, it can be further divided into three zones. A system from zone $1$ experiences only one second-order voltage-induced phase transition from the ordered into the disordered phase, \cf \fig{fig:philic} (note that since our system is symmetric, there are in fact two transitions, at a positive potential, $u_1$, and at a negative potential $-u_1$; the same concerns all zones, hence hereafter we focus only on positive $u$). Zone $2$ contains two transitions (for $u>0$). This is because the transition surface makes a zigzag for increasing voltage (\fig{fig:pd}a,c), and the system first undergoes a transition from the disordered to the ordered phase and then back to the disordered phase, \cf \fig{fig:phobic} (we recall that, at sufficiently high potentials, the system is in the disordered phase, independently of $I$ and $w$). Finally, in zone $3$, there are three transitions. One is either of first or second-order from the ordered to the disordered phase, and then two second-order transitions to the ordered and back to the disordered phase (see \fig{fig:pd}b-c, \cf \fig{fig:zone3}). 

    \end{enumerate}

\Fig{fig:pd}c shows all four zones in the $(w, u)$ plane for a constant value of $\beta I = 2.18$; the cuts at constant $w$ are shown in \figs{fig:philic}a, \ref{fig:phobic}a and \ref{fig:zone3}a. We shall discuss below the charging and capacitive characteristics of a superionic liquid in each of these four zones. It will be convenient, however, to discuss them separately for positive and negative transfer energies, which roughly correspond to ionophilic and ionophobic pores, respectively.

\subsection{Positive transfer energies}

\begin{figure}[t]
\begin{center}
	\includegraphics[width=0.9\textwidth]{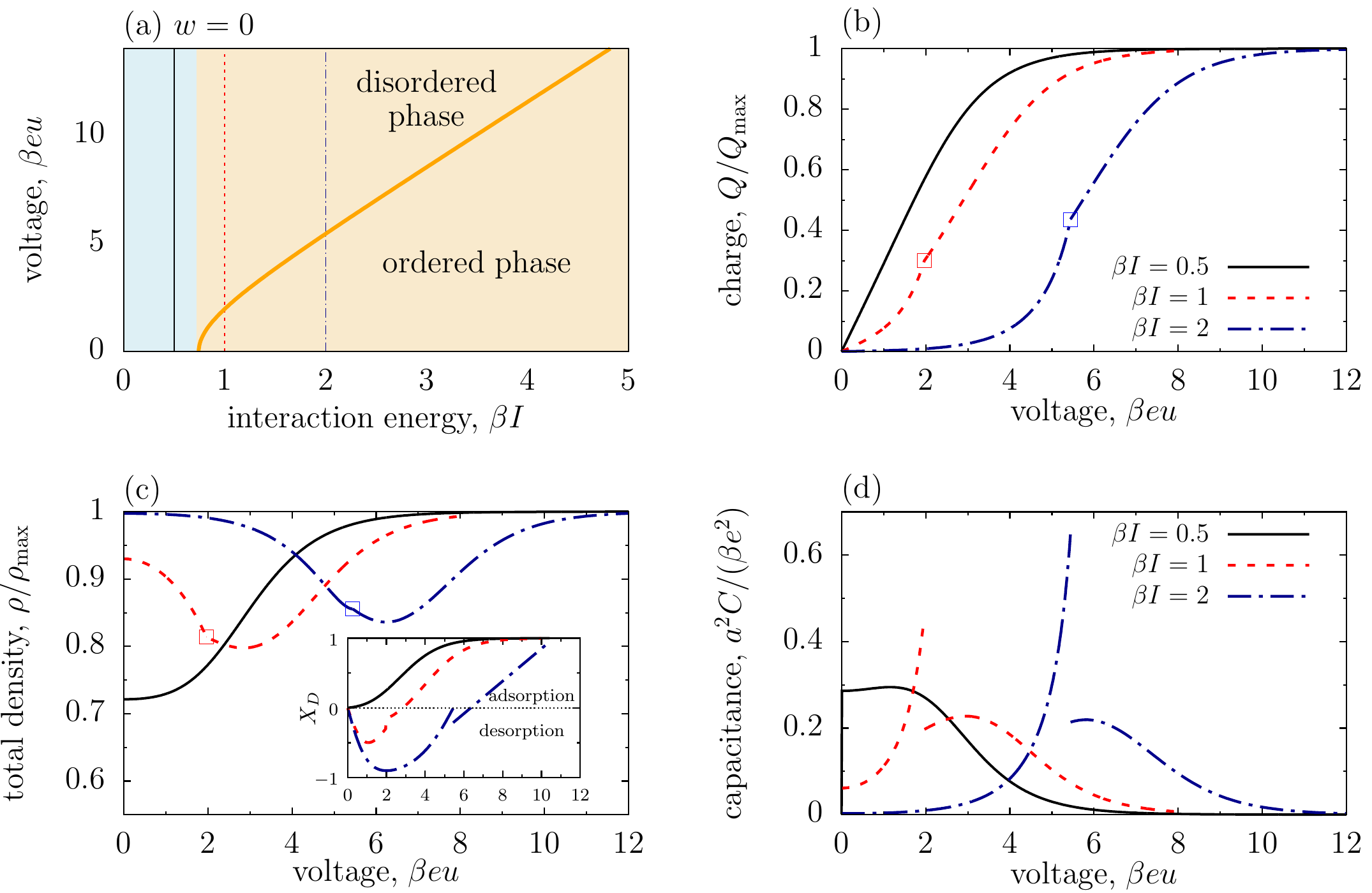}
    \caption{Charging, capacitive and phase behavior of a superionic liquid in ionophilic pores ($w=0$). (a) Phase diagram in the ($I$, $u$) plane, where $I$ is the ion-ion interaction parameter and $u$ the applied voltage. Ordered and disordered phases are separated by a line of second-order phase transitions (solid orange line). The thin vertical lines show the values of $\beta I$ used in (b-d). The light yellow region indicates zone $1$ and the blue domain is zone $0$ (see \fig{fig:pd}b-c). (b) Accumulated charge, (c) total ion density, and (d) differential capacitance as functions of voltage for three chosen values of $\beta I$. The inset in (c) shows the charging parameter, \eq{eq:XD}. The open squares in (b-c) indicate the transition points, where the charge and the total ion density show cusps. Voltage is measured in units of thermal voltage $(\beta e)^{-1}$ ($\approx 26$mV at room temperature) and capacitance is measured in units of thermal electric capacitance~\cite{kondrat:nh:16} ($\beta e^2 \approx 6.2$aF at room temperature) per $a^2$, where $a$ is the lattice constant; for $a=1$nm $\beta e^2/a^2 \approx 620 \mu$F/cm$^2$.
	\label{fig:philic}
}
\end{center}
\end{figure} 

In \fig{fig:philic}a we present the phase diagram in the $(I, u)$ plane for $w=0$, which consists of zones $0$ and $1$. In zone $0$, an ionic liquid is in the disordered phase at any applied potential (blue region in \fig{fig:philic}a). In zone $1$ (light yellow region in \fig{fig:philic}a), it is in the ordered phase at $u=0$, but undergoes a \emph{second-order} phase transition to the disordered phase as the voltage increases. Thus, one can distinguish two types of behavior: (i)~the charge, total ion density and capacitance are all smooth functions of $u$ ($0$-zone, solid lines in \fig{fig:philic}b-c); (ii)~the charge and total ion density exhibit cusps at the transition, whereas the capacitance and the charging parameter $X_D$ show discontinuities (zone $1$, dash and long-dash lines in \fig{fig:philic}b-c).

It is interesting to note that, in zone $0$, charging is driven by swapping co-ions for counter-ions and by counter-ion adsorption, \ie, $X_D > 0$. In zone $1$, however, the system first expels the in-pore co-ions via swapping and co-ion \emph{desorption} ($X_D < 0$), and only then the counter-ion adsorption commences; this point virtually coincides with the transition voltage (the inset in \fig{fig:philic}c). Remarkably, the charge remains nearly constant (and the capacitance is low or vanishes) in the ordered phase before the onset of the transition (dash lines in \fig{fig:philic}b,d). This is likely because changing the ion density (significantly) can distort the ordered phase, which is thermodynamically unfavorable.

It is instructive to compare the charging behavior predicted by our model with the results of other models considered in the literature. In particular, a continuous mean-field model of \mycite{kondrat:jpcm:11} predicted a first-order phase transition between co-ion rich and co-ion deficient phases, with the total ion density and charge experiencing a jump (rather than a cusp) at the transition \cite{kondrat:jpcm:11, lee:prx:16}. However, \emph{charging} behavior was similar in that the counter-ion adsorption starts only after the system expels the co-ions via swapping and/or desorption. A similar charging behavior was observed in molecular dynamics (MD) and Monte Carlo (MC) simulations. \citeauthor{vatamanu:acsnano:15}~\cite{vatamanu:acsnano:15} studied a few ionic liquids by atomistic molecular dynamics simulations and found that charging is often accompanied by a first-order transition, as predicted in \mycite{kondrat:jpcm:11}. Similarly, in a combined MC and MD study, \citeauthor{breitsprecher:jcp:17:mcmd} \cite{breitsprecher:jcp:17:mcmd} also observed that the counter-ion adsorption was preceded by co-ion desorption, although they did not report on any phase transition.

\subsection{Negative transfer energies and multiple re-entrance phenomena}

The region of negative transfer energies mainly corresponds to ionophobic pores, whose existence is yet to be demonstrated experimentally. However, this region includes all four zones and is far more interesting than the region of conventional, ionophilic pores. Since charging in zones $0$ and $1$ appears to be very similar to the case of positive $w$, our primary focus will be on zones $2$ and $3$, although we shall briefly discuss zone $0$ as well.

\subsubsection{Zones $0$ and $2$}

\begin{figure}[t]
\begin{center}
	\includegraphics[width=0.9\textwidth]{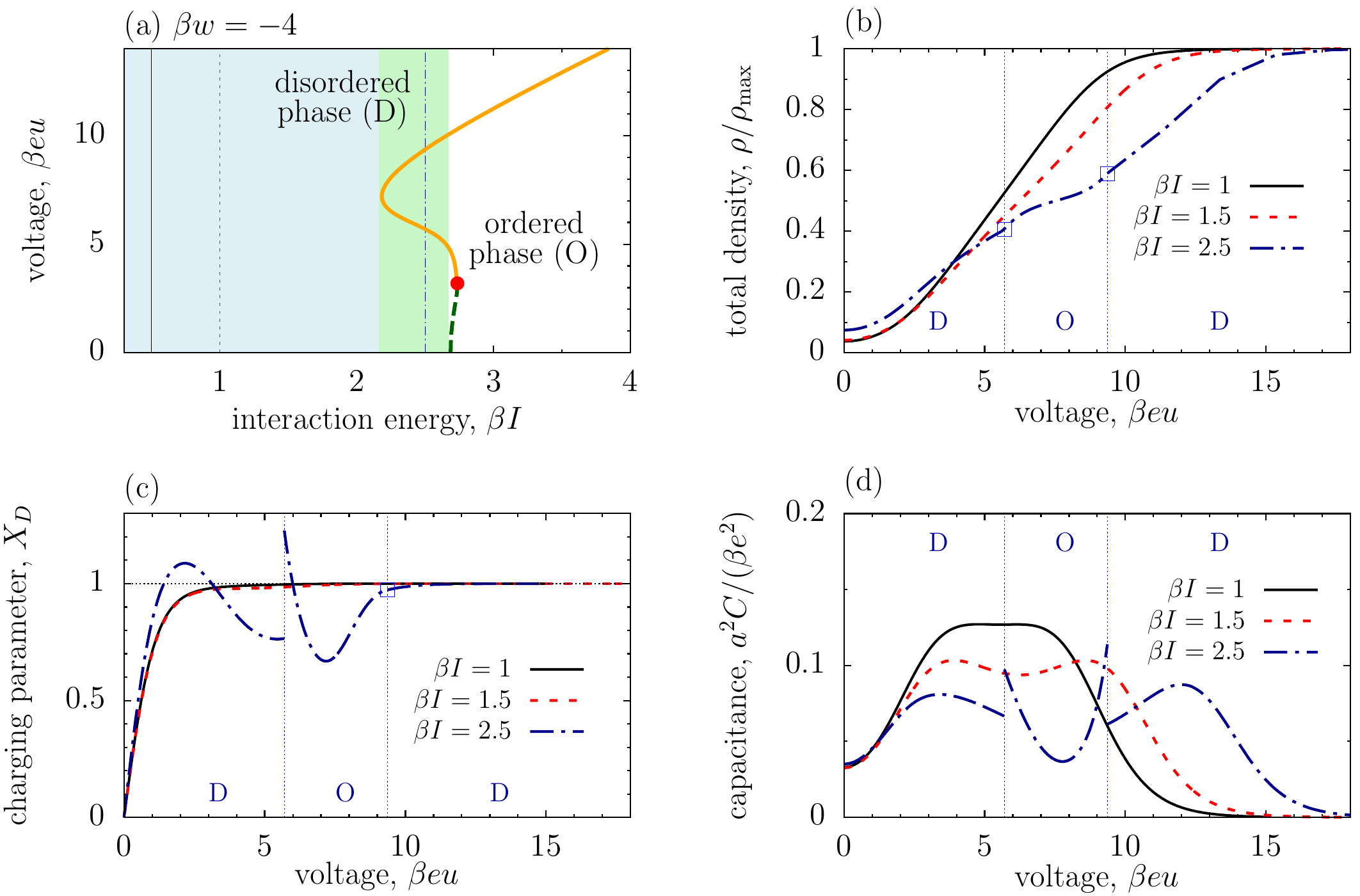}
    \caption{Capacitive, charging and phase behavior of a superionic liquid in ionophobic pores with $\beta w=-4$. (a) Phase diagram in the ($I$, $u$) plane, where $I$ is the ion-ion interaction parameter and $u$ the applied voltage. Ordered (O) and disordered (D) phases are separated by a line of second-order transitions (solid line), and by a line of first-order transitions (dash line), which meet at a tricritical point (full circle). Thin vertical lines show the values of $\beta I$ used in (b-d). This phase diagram contains all four zones, but only $0$-zone (light blue) and zone $2$ (green) are highlighted. (b) Total ion density, (c) differential charging parameter $X_D$ and (d) differential capacitance as functions of voltage for three chosen values of $\beta I$. The open squares and thin vertical lines indicate the location of the transitions for $\beta I = 2.5$. Voltage is measured in units of thermal voltage $(\beta e)^{-1}$ ($\approx 26$mV at room temperature) and capacitance is measured in units of thermal electric capacitance~\cite{kondrat:nh:16} ($\beta e^2 \approx 6.2$aF at room temperature) per $a^2$, where $a$ is the lattice constant; for $a=1$nm $\beta e^2/a^2 \approx 620 \mu$F/cm$^2$.
	\label{fig:phobic}
}
\end{center}
\end{figure} 

\Fig{fig:phobic}a shows the phase diagram in the $(I, u)$ plane for $\beta w=-4$. There are ordered and disordered phases, as before, but here they are separated by lines of first-order (dash line) and second-order (solid line) phase transitions, which meet at a tricricical point (full circle). This phase diagram, in fact, includes all four zones; in this section, however, we shall focus on zones $0$ and $2$ only. To study them, we have chosen the values of $\beta I$ such that the system is in the disordered phase at no applied voltage (thin vertical lines in \fig{fig:phobic}a). For these values of $\beta I$, the pores are (almost) empty at $u=0$ (\fig{fig:phobic}b).

For $\beta I$ corresponding to $0$-zone (solid and dash lines in \fig{fig:phobic}), no phase transition occurs, and the total ion density, charge, charging parameter $X_D$ and capacitance are all continuous functions of $u$, similarly as in the $0$-zone for $w>0$ (the solid lines in \fig{fig:philic}b-d). Here, however, the capacitance shows a \emph{transformation} (not a transition) between one and two-peak shapes, which we had not observed for positive $w$. This behavior can be understood as follows. At small voltages, the system has to overcome an energy barrier for the ions to enter a pore; the potential at which this happens corresponds to the first peak. The second peak appears above a threshold voltage at which the system overcomes an unfavorable interaction between the ions of the same sign; this threshold voltage decreases for decreasing $I$, so that the first and the second peaks merge into a single peak at low values of $I$. Clearly, no energy barrier exists for ions to enter an ionophilic pore, hence only one peak is observed in that case (the solid line in \fig{fig:philic}d).

In zone $2$, the capacitance shows a similar two-peak behavior (blue long-dash line in \fig{fig:phobic}d). In this case, however, charging is discontinuous at \emph{two} second-order phase transitions, as the system experiences a transition from the disordered to the ordered phase and then \emph{re-enters} the disordered phase for increasing voltage. After the onset of the first transition, both counter and co-ions enter a pore, in response to the applied potential, as evidenced by the charging parameter $X_D$, which becomes larger than unity (\fig{fig:phobic}c). At the transition voltage, the system overcomes an energy barrier for ions to enter the pore (similarly as for $\beta I = 1.5$, see the red short-dash lines in \fig{fig:phobic}), and both sorts of ions are adsorbed into the pore, likely in order for the ions to be able to form an `antiferromagnetic' type of ordering. Interestingly, the total ion density, and hence the charge, do not change appreciably within the ordered phase, which is possible because the excess ions would distort the ordered phase. This in fact occurs at the second transition, also denoted by the squares in \fig{fig:philic}b,c (at $\beta e u \approx 9.4$).

\subsubsection{Zone 3}

\begin{figure}[t]
\begin{center}
	\includegraphics[width=0.9\textwidth]{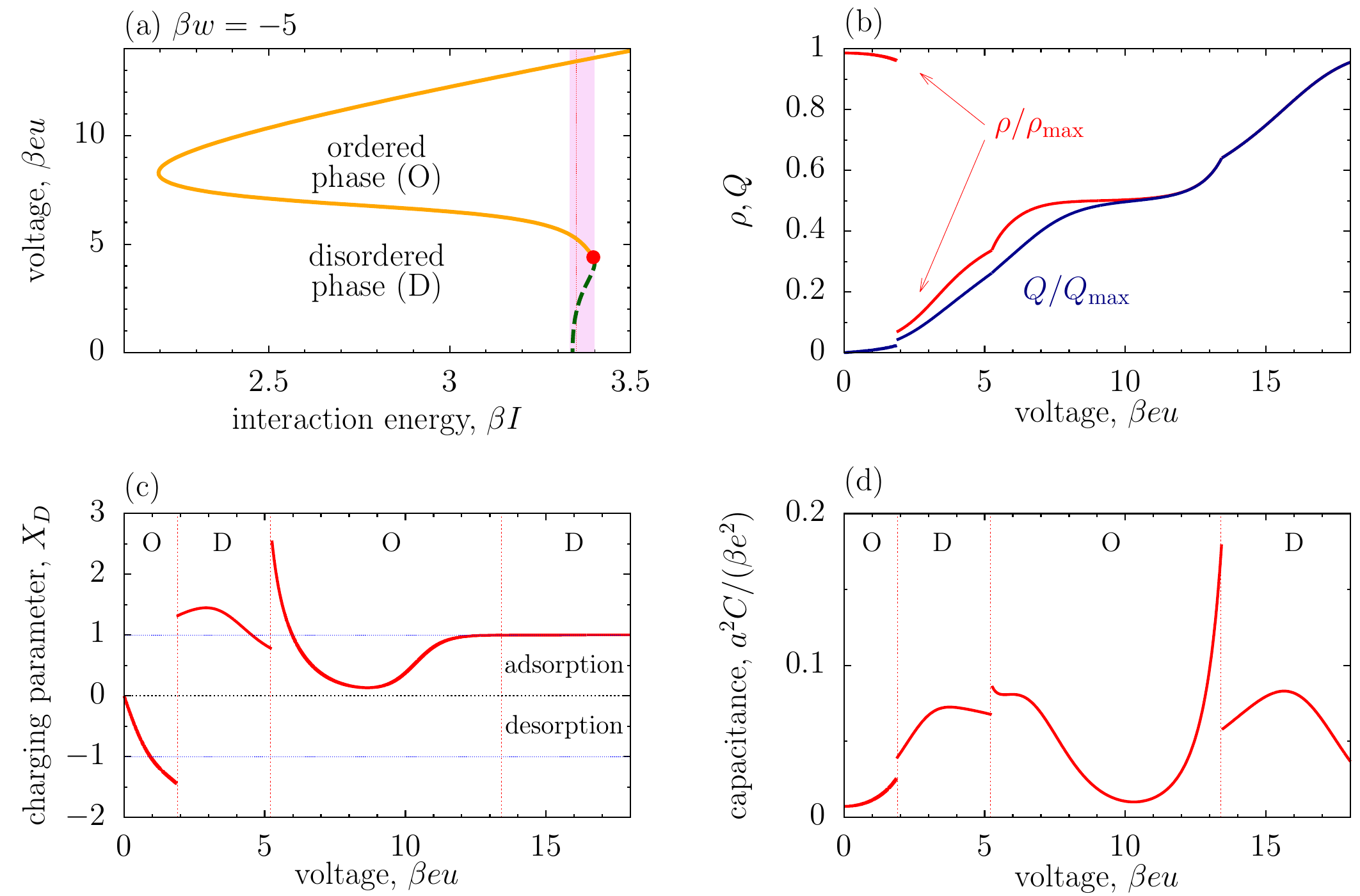}
    \caption{Capacitive, charging and phase behavior of a superionic liquid in zone 3 with $\beta w=-5$. (a) Phase diagram in the ($I$, $u$) plane, where $I$ is the ion-ion interaction parameter and $u$ the applied voltage. Ordered (O) and disordered (D) phases are separated by a line of second-order transitions (solid line), and by a line of first-order transitions (dash line), which meet at a tricritical point (full circle). Thin vertical line shows the value of $\beta I = 3.25$ used in (b-d). The shaded region denotes zone $3$. (b) Accumulated charge and total ion density, (c) differential charging parameter $X_D$ and (d) differential capacitance as functions of voltage for $\beta I = 3.35$. Thin vertical lines denote the locations of the transitions. Voltage is measured in units of thermal voltage $(\beta e)^{-1}$ ($\approx 26$mV at room temperature) and capacitance is measured in units of thermal electric capacitance~\cite{kondrat:nh:16} ($\beta e^2 \approx 6.2$aF at room temperature) per $a^2$, where $a$ is the lattice constant; for $a=1$nm one has $\beta e^2/a^2 \approx 620 \mu$F/cm$^2$.
	\label{fig:zone3}
}
\end{center}
\end{figure} 

We first note that in zone $3$ (and in zone $1$, not shown), despite the negative transfer energies ($w<0$), a pore is nearly fully occupied by ions at $u=0$ (\fig{fig:zone3}b). This is because of the strong interaction between the ions of the opposite sign (\ie, large $\beta I$), which promotes the formation of an ordered phase, characterized by a high density. However, as the voltage increases, \emph{both} types of ions are expelled from the pore, as manifested by \emph{negative} $X_D$ (\fig{fig:zone3}c), so that the pore becomes effectively ionophobic. This process is characterized by a drastic jump in the total ion density (\fig{fig:zone3}b), and by a delta-like peak in the charging parameter ($X_D \to -\infty$ at the transition, which occurs at $\beta e u \approx 1.9$, \fig{fig:zone3}c, not shown for clarity).

After having expelled both co- and counter-ions, the charging proceeds via \emph{co-adsorption} of cations and anions, \ie, both types of ions are adsorbed into the pore again, but in a different proportion. This process is amplified when the system \emph{re-enters} the ordered phase (note high $X_D > 1$ above the transition at $\beta e u \approx 5.2$, \fig{fig:zone3}c). This is likely because more co-ions are required to form an ordered phase than present in the pore at the transition. As the voltage increases further, these excess co-ions are removed from the pore via swapping ($0 < X_D <1$), and eventually the system \emph{re-enters} the final, disordered phase via another second-order phase transition.

This charging behavior manifests itself in the voltage dependence of the differential capacitance, which exhibits three jumps in the course of charging (\fig{fig:zone3}d). It is particularly interesting to point out a relatively low capacitance in the second-ordered phase around $\beta eu = 10$, where the main charging mechanism is swapping ($X_D \approx 0$, \fig{fig:zone3}c). In contrast, above the third transition (which is at $\beta eu \approx 13.3$), the charging is solely due to counter-ion adsorption and is characterized by a relatively high capacitance. This might seem to be in disagreement with the reasoning presented in \mycite{kondrat:nh:16}, where adsorption has been shown to provide the lowest capacitance, and swapping and desorption the highest (this is due to an additional entropic costs associated with the counter-ion adsorption). In the present case, however, swapping occurs within the (`antiferromagnetically') ordered phase, and hence the system has to overcome an additional barrier of breaking this order, which leads to low capacitances (such ordering effects have not been taken into account in \mycite{kondrat:nh:16}).

Thus, zone $3$ appears to be a spectacular region, showing remarkably rich voltage-induced phase and capacitive behavior. We note, however, that it lies in a very narrow domain of $I$ and $w$ and it might be difficult to detect it in experiments or by simulations.

\section{Conclusions}
\label{sec:concl}

We have presented a simple model, \eq{eq:H}, for ionic liquids (ILs) confined to narrow slit-shaped metallic nanopores of nanostructured electrodes, termed as superionic liquids. The model was solved \emph{exactly} on a Bethe lattice with coordination number $q=3$ (\fig{fig:model}). The obtained solutions allowed us to calculate the complete phase diagram of superionic liquids in the space of ion-ion interaction energy, the energy of transfer of ions from a nanopore into bulk electrolyte, and the potential applied at the pore walls with respect to the bulk. The phase diagram consists of ordered and disordered phases (\fig{fig:model}b,c), separated by surfaces of first and second-order phase transitions, and a line of tricritical points (\fig{fig:pd}). 

Most interesting and relevant to supercapacitors is the behavior of confined ILs with applied potential. We have identified four zones, corresponding to the number of phase transitions experienced by an IL for increasing voltage (called zones 0 to 3). While for conventional ionophilic pores, we found either no transition (zone 0), or a single second-order order-disorder transition (zone 1, see \fig{fig:philic}), the phase behavior in ionophobic pores turned out to be far more interesting. In this case, charging can also proceed smoothly without transition; however, in some range of parameters, we revealed the emergence of two types of \emph{re-entrant} behavior. In zone 2, the system is in the disordered state at zero potential, and undergoes a second-order phase transition to the ordered phase, as voltage increases; then there is a second (second-order) phase transition to the disordered phase at a higher voltage (\fig{fig:phobic}). In zone 3, the behavior is even richer as the system undergoes three transitions: From the ordered to disordered phase, then back to the ordered one, and finally to the disordered phase at a high potential (\fig{fig:zone3}).

This variety of phase transitions is reflected in the charging behavior. For ionophilic pores, we found that the capacitance can be either a continuous function of voltage, or experience a divergency in the case of the second-order transition (\fig{fig:philic}d). For ionophobic pores, charging is again more interesting. We revealed a \emph{transformation} (not a transition) between the shapes with one and two peaks in the capacitance-voltage curves, which can be linked to the ionophobic nature of the pores, and appears because the ions have to overcome an energy barrier due to ionophobicity and another one related to the unfavorable interactions between co-ions \footnote[4]{A similar one-to-two peak transformation in the capacitance-voltage shape has been observed in cylindrical pores, see \mycite{lee:prl:14}}. In the case when the system experiences two second-order transitions (zone 2), there are additionally two singularities in the capacitance (\fig{fig:phobic}d); in zone 3, there are three such  singularities (\fig{fig:zone3}d). 

It is also interesting to note that charging of ionophobic pores often proceeds in a manner in which both types of ions are adsorbed into or removed from a pore. This is manifested by the behavior of the charging parameter, which becomes larger than unity or smaller than $-1$, respectively (see \eq{eq:XD} and \figs{fig:phobic}c and \ref{fig:zone3}c). 

    Our findings may have important practical implications. It is known that ordered phases typically exhibit sluggish dynamics, particularly ordered (or quasi-ordered) ionic liquids in slit nanopores \cite{kondrat:nm:14, he15a, Breitsprecher2018}. The knowledge of the parameter space corresponding to such phases may guide to avoid a potential slowdown of charging. Clearly, the precise location of the ordered phases in real systems will be affected by many factors neglected in our model, such as ion-size asymmetry, pore-wall roughness, ion polarizability, \etc (see \sect{sec:limits}). However, as most analytical models, it may serve as a reference point for further and more detailed investigations of superionic liquids, and in particular of the influence of ordering and phase transitions on charging dynamics.

    To solve our model analytically, we have used the Bethe-lattice approach, which assumes that the ions reside on the sites of a Bethe lattice. In this formulation, the partition function, and hence all charging characteristics, can be evaluated \emph{exactly}. We thus expect that our theory captures properly the essential physics of ionic liquids in conducting slit nanoconfinements. However, as we discussed in \sect{sec:limits}, our model neglects a number of important aspects of \emph{real} nanoporous electrodes. In particular, we assumed that an electrode consists of mono-disperse perfectly conducting ultra-narrow slit pores, while typical electrodes are often random porous media fabricated from carbon. This might make the experimental validation of our predictions currently challenging. Nevertheless, the recent progress in low-dimensional porous materials with well-defined pore structure \cite{lukatskaya13a, Naguib2013, Zhao2014, liu:nanolett:10, zhu, tsai:ne:13, chen:jpcl:graphene:13} seems promising in developing the electrodes suitable for our predictions to be tested. We therefore hope that the theory presented in this work will guide the future experiments, providing new physical insights into the charge storage in narrow nanoconfinements.

\begin{acknowledgements}
    A.A.K.~thanks Humboldt foundation for supporting his visit to Max-Planck Institute for Intelligent Systems (MPI-IS, Stuttgart). A.A.K. and S.K. are grateful to Professor Dietrich for hospitality during their simultaneous stay at MPI-IS, when part of this work has been done, and for fruitful discussions. M.D. was supported in part by the ERC grant No. FPTOpt-277998 at the initial stage of his work. M.D. acknowledges the hospitality of LPTMC of University Paris 6 (Sorbonne University)  where the essential part of this work has been accomplished.
\end{acknowledgements}

\appendix

\section{Details of calculations}
\label{app:calcs}

Since the Bethe-lattice calculations for non-zero voltage, $u \not=0$, are similar to the case $u=0$, we describe them only briefly and refer an interested reader to \mycite{dudka:jpcm:16}, where they have been discussed at length.

For a Cayley tree (\fig{fig:model}b,c) with $q$ branches emanating from the root site, the partition function for model \eqref{eq:H} is $\Xi = \lim_{N\to\infty} \Xi_N$, where $N$ is the number of generations in the Cayley tree and 
\begin{equation}
    \label{pfd}
    \Xi_N=g^q_N(0)+z_{+}g^q_N(+)+z_{-}g^q_N(-),
\end{equation}
where $z_{\pm}=\exp \beta h_{\pm}$, with $h_{\pm}$ given by \eq{eq:h}, and $g_N(0)$ and $g_N(\pm)$ are the partition functions of a branch with the root site being vacant and branches with the root sites being occupied by $\pm$ ions, respectively.

Since each branch consists of $q-1$ identical subbranches, one can write the following recursion relations for $g_N$
\begin{eqnarray}\label{recc}
g_N(0)&=&g^{q-1}_{N-1}(0)+z_{+}g^{q-1}_{N-1}(+)+z_{-}g^{q-1}_{N-1}(-),\nonumber\\
g_N(+)&=&g^{q-1}_N(0)+z_{+}e^{-\beta I}g^{q-1}_{N-1}(+)+z_{-}e^{\beta I}g^{q-1}_{N-1}(-),\nonumber\\
g_N(-)&=&g^{q-1}_N(0)+z_{+}e^{\beta I}g^{q-1}_{N-1}(+)+z_{-}e^{-\beta I}g^{q-1}_{N-1}(-).
\end{eqnarray}
By introducing new variables,
\begin{equation}\label{nvar}
x_N=\frac{g_N(+)}{g_N(0)}
\quad 	\textrm{and} \quad
y_N=\frac{g_N(-)}{g_N(0)},
\end{equation}
one gets a system of two coupled recursion relations
\begin{eqnarray}\label{recq}
x_N&=&\frac{1+z_{+}e^{-\beta I}x^{q-1}_{N-1} +z_{-}e^{\beta I}y^{q-1}_{N-1}}{1+z_{+}x^{q-1}_{N-1}+z_{-}y^{q-1}_{N-1}} ,\nonumber\\
y_N&=&\frac{1+z_{+}e^{\beta I}x^{q-1}_{N-1} +z_{-}e^{-\beta I}y^{q-1}_{N-1}}{1+z_{+}x^{q-1}_{N-1}+z_{-}y^{q-1}_{N-1}}.
\end{eqnarray}

In the interior part of the Cayley tree, and in the limit $N\to\infty$, \ie, \emph{on the Bethe lattice} \cite{Baxter}, all sites are equivalent and hence all $\{x_N,y_N\}$ converge to a fixed point or cycle solutions $\{x,y\}$. In this case, excluding the effects of boundary sites by following the procedure of \mycite{Ananikian98} (see also \mycite{Gujrati95}), we have obtained for the \emph{bulk} free energy per site \cite{dudka:jpcm:16}
 \begin{equation}\label{free}
 -\beta f= \frac{q}{2}\ln\left(1+z_{+}x^{q-1}+z_{-}y^{q-1}\right)-
  \frac{q-2}{2}\ln\left(1+z_{+}x^{q}+z_{-}y^{q}\right).
 \end{equation}
To take into account the bipartitte nature of the Caylee tree, we modified our description to include \emph{sublattices} $A$ and $B$. Then, instead of \eq{recq}, one has, taking additionally the limit $N\to\infty$
\begin{align}
    x_A=\frac{1+e^w (e^{- I+ e u}x_B^{q-1} +e^{ I- e u} y_B^{q-1})}{1+e^w(e^{e u}x_B^{q-1}+e^{- e u}y_B^{q-1})}, \nonumber\\
    y_A=\frac{1+e^w (e^{ I+e u}x_B^{q-1} +e^{ -I-e u} y_B^{q-1})}{1+e^w(e^{e u}x_B^{q-1}+e^{- e u}y_B^{q-1})}, \label{A}\\
    x_B = \frac{1+e^w (e^{- I+eu}x_A^{q-1} +e^{ I-eu} y_A^{q-1})}{1+e^w(e^{e u}x_B^{q-1}+e^{- e u}y_B^{q-1})}, \nonumber\\
    y_B = \frac{1+e^w (e^{ I+eu}x_A^{q-1} +e^{-I-eu} y_A^{q-1})}{1+e^w(e^{e u}x_B^{q-1}+e^{- e u}y_B^{q-1})}\label{B},
\end{align}
where we have taken into account the explicit expression for $z_{\pm}$ and assumed $w_+=w_-=w$. The free energy of full system is $[f(x_A,y_A)+f(x_B,y_B)]/2$, with expression for $f(x,y)$ given by \eq{free}. To find $x_A$, $y_A$, $x_B$ and $y_B$, we have solved \eqs{A}{B} with $q=3$ \footnotemark[2] numerically using \verb|Mathematica|.

\subsection{Ordered and disordered phases}
\label{app:phases}

\begin{figure}[t]
\includegraphics[width=.8\textwidth]{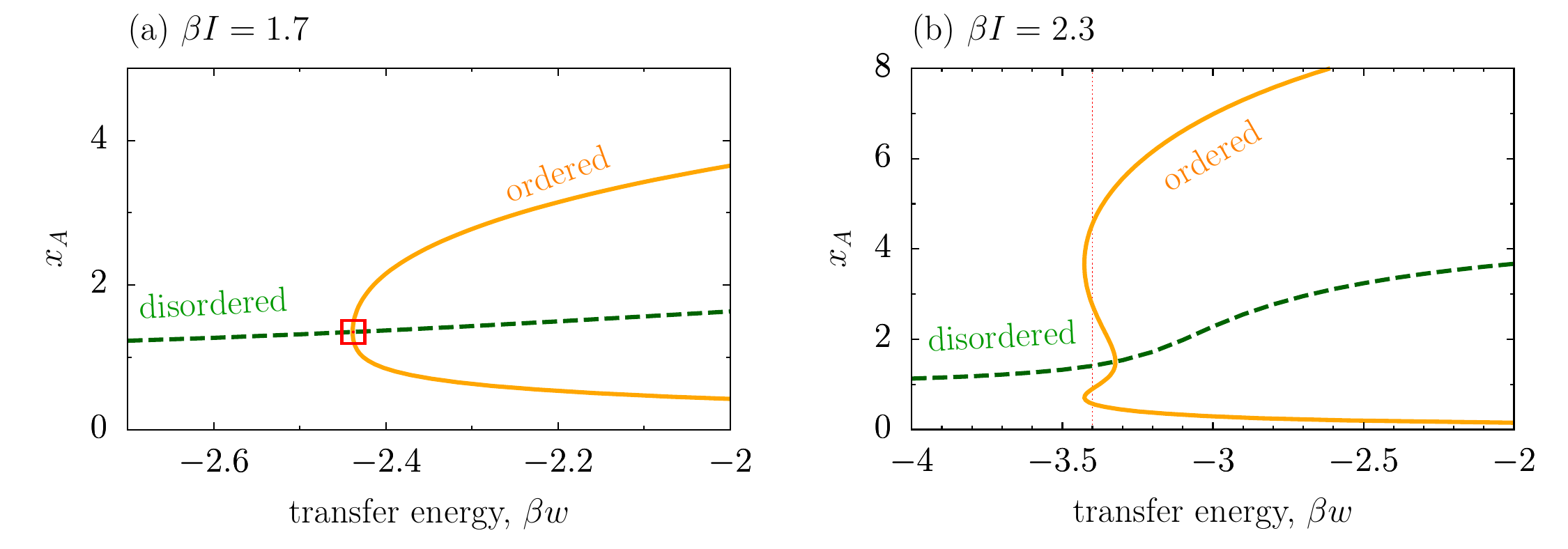}
    \caption{Bethe-lattice solutions for the ordered and disordered phases. (a) Solutions for disordered (dash line) and ordered (solid line) phases for $\beta e u=0.5$ and $\beta I=1.7$. The crossing point between two lines (denoted by the symbol) determines the location of the second-order phase transition. (b) The same but for $\beta I=2.3$, at which the transition is of first order. The location of the transition is determined by comparing the free energies of the ordered and disordered phases (thin vertical line).
}
\label{solutions}
\end{figure}

Similarly as in the case of zero voltage \cite{dudka:jpcm:16}, there are two stable thermodynamic phases for $u \not = 0$, namely:
\begin{enumerate}

    \item \emph{Disordered phase}, described by $x_A=x_B$, $y_A=y_B$ (\fig{fig:model}b). The solution for the disordered phase can be obtained from \eq{A} by putting $x_B=x_A$ and $y_B=y_A$. \Eq{B} is not needed.

    \item \emph{Ordered phase} with $x_A\not=x_B$ and $y_A\not=y_B$ (\fig{fig:model}c). In this case all four equations, \eqs{A}{B}, must be solved. 

\end{enumerate}

Examples of the Bethe-lattice solution for $q=3$ and for $\beta eu=0.5$ and two values of $I$ are shown in \fig{solutions}. \Fig{solutions}a shows $x_A$ as a function of the transfer energy $\beta w$ for $\beta I=1.7$, at which there is a \emph{second-order} phase transition between the ordered and disordered phases. For low values of $w$, there is no ordering in the system (\ie, $x_A=x_B$ and $y_A=y_B$), but with increasing $w$ to a critical value $\beta w \approx -2.44$ (the crossing point between the two lines in \fig{solutions}a, denoted by the symbol), another solution emerges with $x_A\ne x_B$ and $y_A \ne y_B$, which corresponds to the ordered phase (the solution corresponding to the disordered phase becomes unstable).

\Fig{solutions}b illustrates the case of a first-order phase transition ($\beta I = 2.3$). In the vicinity of the crossing point between two lines in \fig{solutions}b, the solution for the ordered phase is \emph{multivalued}. The value of $w$ at which the transition occurs can be found by matching the free energies (\ref{free}) of the disordered and ordered phases ($\beta w\approx -3.4$, denoted by the thin vertical line in \fig{solutions}b).

The above described procedure has been used to calculate the phase diagrams in \figs{fig:pd}, \ref{fig:philic}, \ref{fig:phobic} and \ref{fig:zone3}.

\subsection{Thermodynamic quantities and charging characteristics}

Average densities of $\pm$ ions on the root (central) site of the Cayley tree can be obtained via the variables $x_N$ and $y_N$ as follows \cite{dudka:jpcm:16}
\begin{equation}\label{dense}
\rho_{0,+}=\frac{z_{+}x^{q}_N}{1+z_{+}x^{q}_N+z_{-}y^{q}_N},\quad
\rho_{0,-}=\frac{z_{-}y^{q}_N}{1+z_{+}x^{q}_N+z_{-}y^{q}_N} \,.
\end{equation}
Assuming that the central site belongs to sublattice $A$, \eq{dense} gives for the average density of ions on sublattice $A$ (in the limit $N\to\infty$)
\begin{equation}\label{denseA}
\rho^{A}_{+}=\frac{z_+ x^{3}_{A}}{1+z_+ x^{3}_{A}+z_- y^{3}_{A}},\quad
\rho^{A}_{-}=\frac{z_- y^{3}_{A}}{1+z_+ x^{3}_{A}+z_- y^{3}_{A}}.
\end{equation}
If the central site belongs to sublattice $B$, the expression for the average density can be obtained from \eq{denseA} by interchanging the indices $A \leftrightarrow B$.

The total ion density and the charge density (in units of the elementary charge $e$) on sublattice $A$ are
\begin{equation}
\label{eq:op:def}
\rho^A=(\rho^A_{+}+\rho^A_{-}), \qquad
\delta\rho^A=(\rho^A_{+}-\rho^A_{-}),
\end{equation}
and similarly for sublattice $B$. For the whole system, one obviously has
\begin{equation}\label{totalvar}
\rho=\frac{\rho^{A}+\rho^{B}}{2}, \quad
\delta\rho=\frac{\delta\rho^{A}+\delta\rho^{B}}{2}.
\end{equation}

The charge accumulated in a nanopore is (see \eq{eq:Q})
\begin{equation}\label{g-charge}
Q=e\delta\rho,
\end{equation}
and hence the differential capacitance (see \eq{eq:cap})
\begin{equation}\label{g-capacitance}
C= \frac{ d Q}{d u}= e \frac{ d \delta \rho}{d u}
\end{equation}
and the charging parameter (see \eq{eq:XD})
\begin{equation}\label{xd}
X_D=\frac{e}{C} \frac{ d \rho}{d u}.
\end{equation}
To calculate these quantities analytically, it is convenient to rewrite \eqs{A}{B} as
\begin{eqnarray}
x^2_{A}z_+ &=&\frac{(1-\lambda)(1-x_{B})+\lambda (1-y_{B})}{x_{B}+y_{B}-2 \cosh(\beta I)} \nonumber\\
y^2_{A} z_-&=&\frac{(1-\lambda)(1-y_{B})+\lambda (1-x_{B})}{x_{B}+y_{B}-2 \cosh(\beta I)},  \label{xysubs}\\
x^2_{B}z_+ &=&\frac{(1-\lambda)(1-x_{A})+\lambda (1-y_{A})}{x_{A}+y_{A}-2 \cosh(\beta I)} \nonumber\\
y^2_{B} z_-&=&\frac{(1-\lambda)(1-y_{A})+\lambda (1-x_{A})}{x_{A}+y_{A}-2 \cosh(\beta I)},  \label{xysubs2}
\end{eqnarray}
where $\lambda={e^{\beta I}}/(1+e^{\beta I})$. Plugging these expressions into \eq{totalvar}, one finds
\begin{align}
\delta\rho=1/2\frac{x_A+x_B-y_A-y_B+2(\lambda-1)(x_A x_B-y_A y_B)}{x_A+x_B+y_A+y_B+(\lambda-1)(x_A x_B+y_A y_B)-\lambda(x_A y_B+x_B y_A)-2\cosh(\beta I)}
\end{align}
and
\begin{align}
\rho=1/2\frac{x_A+x_B+y_A+y_B+2(\lambda-1)(x_A x_B+y_A y_B)-2\lambda(x_A y_B+x_B y_A)}{x_A+x_B+y_A+y_B+(\lambda-1)(x_A x_B+y_A y_B)-\lambda(x_A y_B+x_B y_A)-2\cosh(\beta I)}.
\end{align}
To calculate the differential capacitance, \eq{g-capacitance}, and charging parameter, \eq{xd}, one needs the derivatives $d x_{A}/{du}$, ${d y_{A}}/{du}$,  ${d x_{B}}/{du}$, ${d y_{B}}/{du}$. They can be obtained by differentiating both side of \eqs{xysubs}{xysubs2} with respect to $u$, and then solving the resulting system of equations for the corresponding derivatives numerically.

\bibliography{bethe,supercaps}

\end{document}